\documentstyle [11pt]{article}
\def \d{\partial}

\def \be{\begin{equation}}
\def \ee{\end{equation}}

\def \ra{\rangle}
\def \la{\langle}
\def \1{{\bf 1}}
\def \a{\alpha}
\def \b{\beta}

\def \fb{\bar{f}}
\def \fh{\hat{f}}
\def \gb{\bar{g}}
\def \gh{\hat{g}}
\def \Tr{\mbox{Tr}}
\def \fr{\frac}

\def \S{{\cal{S}}}
\def \Dt{\Delta\tau}
\begin{document}
\title{Improved Langevin Methods for Spin Systems}
\author{I T Drummond and R R Horgan  \\
        Department of Applied Mathematics and Theoretical Physics \\
        University of Cambridge \\
        Silver St \\
        Cambridge, England CB3 9EW}
\maketitle
\begin{abstract}

We investigate methods for variance reduction and the elimination of systematic
error in a Fourier accelerated Langevin scheme for general spin models. We
present
results for the $SU(3)\times SU(3)/SU(3)$ model in two-dimensions that are
consistent with those
from multi-grid methods. We argue that the timing for the Langevin method makes
it comparable to multi-grid for a given level of error.
\end{abstract}
\vfill
DAMTP-92-72
\pagebreak
\section{Introduction}
Langevin simulation methods may be applied to any continuous spin model
\cite{us}.
While certain models of the $O(N)$ type may respond best to cluster algorithms
\cite{Wlf}
there is no comparable method for those of $SU(N)$ or $CP(N)$ type. These have
been
tackled by multi-grid methods with good results \cite{HasMey1,HasMey2,Edw}. We
show that when the
appropriate variance reduction techniques and extrapolation methods for the
removal of
systematic errors have been applied, the Langevin scheme can compete on a
reasonably
equal footing with the multi-grid approach at least for the case $N=3$~. Higher
values of $N$ remain to be investigated. A preliminary version of this work was
reported
at Lattice91 \cite{itdrrh}

\section{Langevin Scheme}

For the purposes of exposition we will assume that we are dealing with an
$SU(N)\times SU(N)/SU(N)$
model on an $L^D$ hyper-cubical lattice. We will report results with $N=3$ and
$D=2$ for various values of $L$~.
The action is
\be
A(S)=\fr{g}{2}\sum_{n,\mu}\left(\Tr(S_{n+\mu}^{\dag}S_n)+\mbox{c.c.}\right)~~,
\ee
where $S_n$ is an element of $SU(N)$ associated with site $n$ and $\mu$ runs
over forward pointing
nearest neighbour sites. The probability distribution we wish to simulate is
$P_0(S)\propto\exp(A(S))~$.
This distribution is the stationary solution of the Fokker-Planck equation
\be
\fr{\d}{\d\tau}P(S,\tau)=HP(S,\tau)~~,
\ee
where
\be
H=\sum_{nma}\kappa_{nm}D^a_n\left(D^a_m-U^a_m\right)~~.
\ee
Here $D^a_n$ is the covariant derivative appropriate to site $n$ in group
direction $a$ defined as follows \cite{DDH}.
Let $\{\Lambda^a\}$ be a basis of generators for the defining representation of
$SU(N)$~.
They satisfy the Lie algebra relation $[\Lambda^a,\Lambda^b]=C_{abc}\Lambda^c$
where the $\{C_{abc}\}$
are the structure constants of the group. Let $e^{\epsilon.\Lambda}S$ stand for
the set of
spins $\{e^{\epsilon_n^a\Lambda^a}S_n\}$~. Then
\be
D^a_nf(S)=\fr{\d}{\d\epsilon_n^a}f(e^{\epsilon.\Lambda}S)
\left.\right|_{\epsilon=0}~~.
\ee
The covariant derivatives satisfy the Lie algebra relation
$[D^a_n,D^b_m]=-\delta_{nm}C_{abc}D_n^c~~$.
The drift term in the Fokker-Planck equation is $U^a_n=D^a_nA(S)$~. Clearly for
any choice of
the ``diffusivity'' tensor $\kappa_{nm}$ we have $HP_0(S)=0$~.

The presence of $\kappa_{nm}$ allows us to exploit the idea of Fourier
acceleration \cite{Bat}.
In order to preserve translation invariance $\kappa_{nm}$ depends only on the
separation $n-m$
of the two sites.
The precise form of the its functional depencence is fixed by specifying the
shape of its Fourier
transform. Guided by ideas of free field theory we choose \cite{Bat}
\be
\tilde{\kappa}(k)=\fr{(L(k)+M^2)_{\mbox{max}}}{L(k)+M^2}~~,
\ee
where $L(k)$ is the lattice Laplacian and $M$ is referred to as the
acceleration parameter and is
in practice adjusted to the expected value of the correlation length under
investigation.

The hermitian conjugate operator of $H$ which we will use below is
\be
H^{\dag}=\sum_{nma}\kappa_{nm}\left(D^a_n+U^a_n\right)D^a_m~~.
\ee

We set the common diagonal value of the diffusivity tensor $\kappa_{nn}=K$~.
The simulation was performed using a two-stage Fourier accelerated Runge-Kutta
algorithm presented in ref \cite{us} in a form appropriate to $O(N)$ groups.
For convenience we give the algorithm here in a more general notation. The
updating
step proceeds by first constructing an intermediate position in configuration
space.
\be
S_n^{(\mbox{old})}\rightarrow
S_n^{(1)}=e^{\epsilon^{(1)a}_n\Lambda^a}S_n^{(\mbox{old})}~~,
\ee
where
\be
\epsilon^{(1)a}=\fr{1}{2}A_{nm}U^a_m\Delta\tau+\sqrt{\Delta\tau}
\left(B^{\fr{1}{2}}\right)_{nm}\eta^{(1)a}_m~~.
\ee
The matrices $A$ and $B$ are
\be
A_{nm}=\kappa_{nm}+\Delta\tau
C\left[\fr{1}{4}(\kappa_{nm})^2-\fr{1}{6}K\kappa_{nm}\right]~~,
\ee
and
\be
B_{nm}=\kappa_{nm}+\Delta\tau
C\left[\fr{1}{4}(\kappa_{nm})^2-\fr{1}{3}K\kappa_{nm}\right]~~.
\ee
The final step to the updated configuration is
\be
S_n^{(\mbox{old})}\rightarrow
S_n^{(\mbox{new})}=e^{\epsilon^{(2)a}_n\Lambda^a}S_n^{(\mbox{old})}~~,
\ee
where
\be
\epsilon^{(2)a}=A_{nm}U^{(1)a}_m\Delta\tau
+\sqrt{\Delta\tau}\left(B^{\fr{1}{2}}\right)_{nm}
(\eta^{(1)a}_m+\eta^{(2)a}_ma)~~,
\ee
the set of quantities $\{\eta^{(1)a}_m\}$ and $\{\eta^{(2)a}_m\}$ being
independent gaussian random
numbers with zero mean and unit variance. The superfix $(1)$ on the drift term
indicates it has been
evaluated at the intermediate configuration $\{S^{(1)}_n\}~$.

\section{Variance Reduction}
The method of modifying an estimator so that its mean is unchanged but its
variance is reduced has been
commonly exploited. In the present case we use the structure of the Langevin
process itself to create
such variance reduced estimators. The technique is based on the result that the
probability distribution $P_0(S)$
we use for averaging is the null eigenfunction of the operator $H$ in the
Fokker-Planck equation.
We have
\be
HP_r(S)=E_rP_r(S)~~,
\ee
where $E_r<0,~~r\geq 1$ and $E_0=0$~. We normalise $P_0(S)$ so that
\be
\la f(S)\ra=\int d\mu(S)P_0(S)f(S)~~,
\ee
where $d\mu(S)$ is the group invariant measure on the manifold of spins.
Because $HP_0(S)=0$
it follows after an appropriate integration by parts that
\be
\la f(S)\ra=\int d\mu(S)P_0(S)W(H^{\dag})f(S)~~,
\ee
where $W(z)$ is any power series in $z$ such that $W(0)=1$~. This implies that
we can use
$f'(S)=W(H^{\dag})f(S)$ as an estimator for the same quantity as $f(S)$. There
is then the
possibility that $f'(S)$ has a smaller variance than $f(S)$ itself.

The set $\{\psi_r(S)=P_r(S)/P_0(S)\}$ are the eigenfunctions of $H^{\dag}$
\be
H^{\dag}\psi_r(S)=E_r\psi_r(S)~~.
\ee
They satisfy the orthogonality relations
\be
\int d\mu(S)P_r(S)\psi_s(S)=\delta_{rs}~~.
\ee
Clearly $\psi_0(S)\equiv 1$~. If we expand $f(S)$ on the set $\{\psi_r(S)\}$ we
obtain
\be
f(S)=\la f(S)\ra+\sum_{r=1}^{\infty}\a_r\psi_r(S)~~,
\ee
for some set of coeficients $\{\a_r\}$~. The new estimator $f'(S)$ is then
\be
f'(S)=\la f(S)\ra+\sum_{r=1}^{\infty}\a_rW(E_r)\psi_r(S)~~,
\ee
and its variance is
\be
\sigma^2(f')=\sum_{r=1}^{\infty}(\a_rW(E_r))^2~~.
\ee
When $W(z)\equiv 1$, $\sigma^2(f')$ reduces to $\sigma^2(f)$~. If now we
minimise $\sigma^2(f')$
with respect to the parameters in $W(z)$ we can expect to find
$\sigma^2(f')<\sigma^2(f)$~.
A simple choice which we implement here is $W(z)=1+bz$~. Strategies for
choosing the value
of $b$ are explained in the next section.

\section{Systematic Errors}

The above theory is only applicable to the simulated results on the assumption
that
the simulated probability distribution has no systematic errors. In fact our
Langevin
scheme exhibits systematic errors that are $O(\Delta\tau^2)$~. We deal with
this
problem by a statistically controlled extrapolation proceedure combined with
variance reduction.

In outline the procedure for evaluating the mean of an observable $f(S)$ is to
measure
$\la f\ra\mbox{and} \la\fh\ra$ where $\fh=H^{\dag}f$ together with the
correlators
$\la f^2\ra, \la\fh^2\ra \mbox{and} \la f\fh\ra~$ for a number of runs
with different values of values of $\Dt^2$. For the smallest of these values
$\Dt_1^2$
say, we obtain the value of $b$ that yields the minimum variance for the
estimator $f'=f+b\fh$~.
We have
\be
\sigma^2(f')=\sigma^2(f)+2b\sigma^2(f,\fh)+b^2\sigma^2(\fh)~~,
\ee
and minimum therefore
occurs when
\be
b=-\sigma^2(f,\fh)/\sigma^2(\fh)~~,
\label{sig}
\ee
where $\sigma^2(f,\fh)=\la f\fh\ra-\la f\ra\la\fh\ra$~.

Having established the value of $b$ and hence the precise form of $f'$ in this
way we
compute the $\la f'\ra$ and $\sigma^2(f')$ for the other larger choices of
$\Delta\tau$~.
We then do a least squares fit on the assumption that
\be
\la f'\ra=\left.\la f'\ra\right|_{\Delta\tau=0}+\beta\Delta\tau^2~~.
\ee
That is we minimise $\chi^2$ with respect to $\a$ and $\b$ where
\be
\chi^2=\sum_i\fr{1}{\sigma_i^2}(\bar{f'}_i-\a-\b\Delta\tau^2_i)^2~~.
\ee
Here $i$ labels the run that uses time step $\Delta\tau_i$ and $\fb'_i$ and
$\sigma_i^2$
are the estimates of the mean and variance of $f'$ based on run $i^{\mbox{th}}$
run.
The minimising value of $\a$ yields our {\em unbiased} estimate for $\la f\ra$
and is given
by
\be
\a=\fr{\left(\sum_i\fr{\fb'_i}{\sigma^2_i}\right)
\left(\sum_j\fr{\Dt_j^4}{\sigma_j^2}\right)
-\left(\sum_i\fr{\fb'_i\Dt_i^2}{\sigma^2_i}\right)
\left(\sum_j\fr{\Dt_j^4}{\sigma_j^2}\right)}
{\left(\sum_i\fr{1}{\sigma^2_i}\right)
\left(\sum_j\fr{\Dt_j^4}{\sigma_j^2}\right)
            -\left(\sum_i\fr{\Dt_i^2}{\sigma^2_i}\right)^2}~~.
\ee

The advantage of this proceedure is that it not only
provides an estimate of the statistical error on the observable of interest but
through the
resulting $\chi^2$ for the fit provides a test of the assumption of linearity
in $\Delta\tau^2$
for the systematic error. Our evaluation of the statistical error of our
unbiased estimate of $\la f\ra$
is $\sqrt{\sigma^2(\a)}$ where
\be
\sigma^2(\a)=\sum_i\left(\fr{\d\a}{\d\fb'_i}\right)^2\sigma^2_i~~.
\ee
The detailed treatment of the data involves a systematic re-binning proceedure
that we will
discuss when we consider decorrelation of measurements in the next section.

The correlation length cannot be measured directly as an observable. It is
obtained by
fitting simultaneously the
estimates for a number of observables, namely the values of the correlation
function
$G_t=\la g_t\ra$, where $g_t= \Tr\left(\S_t^{\dag}\S_0\right)\ra$~, $\S_t$
being the
spatial average of the spins $\{S_n\}$
for $n$ in time-slice $t$~. Our strategy in this case is a generalization of
the single operator
case discussed above. For a set of values for $\Dt$ we take measurements
of the observables $g_t$~
and the observables $\gh_t=H^{\dag}g_t$~ forming the correlation matrices $\la
g_tg_{t'}\ra$,
$\la g_t\gh_{t'}\ra$ and $\la \gh_t\gh_{t'}\ra$~. Our method of proceeding is
as follows.
We construct a set of improved operators $\{g'_t\}$ where $g'_t=g_t+b_t\gh_t$~.
The coefficients $\{b_t\}$ are to be chosen ultimately to minimise the
statistical error on
our estimate of the correlation length. For simplicity we worked in fact with a
common value $b$
for all the members of the set $\{b_t\}$~. It is important to note that the
optimal value of $b$
appropriate to the correlation length calculation proved to be different from
the value appropriate to the susceptibility calculation.

We then can compute the
correlation matrix $\Sigma_{tt'}=\la g'_tg'_{t'}\ra-\la g'_t\ra\la g'_{t'}\ra$
for a given $\Dt$~.
We make the assumption that
\be
\la g'_t\ra=\left.\la g'_t\ra\right|_{\Dt=0}+h_t\Dt^2~~,
\ee
and obtain an {\em unbiased} estimate for the Green function $G_t=\la g_t\ra$
from the extrapolation to $\Dt=0$
by minimising $\chi^2$ where
\be
\chi^2=
\sum_{itt'}\Sigma_{itt'}^{-1}\left(\gb_{it}-\a_t-\b\Dt_i^2\right)
                             \left(\gb_{it'}-\a_{t'}-\b_{t'}\Dt_i^2\right)~~,
\ee
where $\gb_{it}$ is the numerical estimate for $G_t$ in the run with
$\Dt=\Dt_i$, and
$\Sigma^{-1}_{itt'}$ is the matrix inverse of $\Sigma_{itt'}$ the corresponding
numerical estimate for $\Sigma_{tt'}$~. The variance matrix for the resulting
$\{\a_t\}$ is
\be
\sigma^2(\a_t,\a_{t'})
=\sum_{it''t'''}\fr{\d\a_t}{\d\gb_{it''}}
\Sigma_{it''t'''}\fr{\d\a_{t'}}{\d\gb_{it'''}}~~.
\ee
We use the $\{\a_t\}$ as the estimates for the $\{G_t\}$ extrapolated to
$\Dt=0$ and the
above variance matrix as the estimate for the extrapolated correlator
$\la g_tg_{t'}\ra-\la g_t\ra\la g_{t'}\ra$~. Again we obtain a value for
$\chi^2$ that measures
the goodness of fit for the extrapolation.

To these values we fit the shape
\be
G_t=A+B\cosh \left((t-L/2)/\xi\right)~~,
\label{formula}
\ee
determining the values
of $A, B~\mbox{and}~\xi$ by minimising
\be
\chi^2=\sum_{tt'}\left(\a_t-G_t\right)
\Sigma^{-1}_{tt'}(a)\left(\a_{t'}-G_{t'}\right)~~,
\ee
where $\Sigma_{tt'}(a)=\sigma^2(\a_t,\a_{t'})$ and $\Sigma^{-1}_{tt'}(a)$ is
its matrix inverse.

The functional form in eq(\ref{formula}) is not appropriate for lowest few
$t$-values
where $G(t)$ is appreciably affected by contributions other than the lowest
state.
We omit these points from the fitting procedure and search for a consistent fit
at
the higher values of $t$ with an acceptable chi$^2$~ of about unity per degree
of freedom.

\section{Singular Value Decomposition of $\Sigma$}

Because $\Sigma_{tt^\prime}$ has some very small eigenvalues it is
possible that there can be some
difficulty in inverting this matrix and hence in obtaining a meaningful
evaluation of
chi$^2$. However, this problem can be circumvented by exploiting the singular
value decomposition (SVD) of $\Sigma$ \cite{Flan} and following
the procedure suggested by Lepage in ref\cite{beth}.

The data for $G_t$ for different values of t are highly positively correlated.
This means that the estimates for $G_t$ lie on a smoother curve than the
individual
errors on each point would suggest. It also means inevitably that there are
eigenvectors of $\Sigma$ associated with very small eigenvalues and that these
eigenvectors have entries which change in sign: a behaviour which is unphysical
and which is not represented in the fitting function eq(\ref{formula}). For a
large lattice
there are many such ``unphysical'' eigenvectors and their associated amplitudes
in
the expansion of $G_t$ on the eigenbasis each contribute to chi$^2$ weighted by
the inverse of a small eigenvalue i.e., a large number. We then
note that the eigenvalues are
only statistical estimates for the true variances, that the number of
unphysical
parameters grows with lattice temporal size and that their contribution to
chi$^2$ is strongly
weighted compared with those of the more physical eigenvectors.

The naive fitting procedure can then produce unsatisfactory results since
chi$^2$ is
dominated by unphysical modes and a moderate change in the
fitting parameters has very little
effect on their contribution to chi$^2$ since the smooth fitting function has
little overlap with the unphysical modes. The consequence is that large
variations in
the parameters occur in an attempt to reduce the bulk contribution to chi$^2$
and the fitted values are
thus controlled not by the physical signal but by statistical fluctuations in
amplitudes known to be zero .
This can result in a fitted function which clearly
does not pass through the smooth set of data points anywhere near as well as it
should.
To remedy the situation the inverse of $\Sigma$ is computed
by first making a singular value decomposition and then inverting the
eigenvalues but
replacing the inverse of all but the largest ones by {\it zero}. This removes
the
effect of unphysical modes and means that we need not add oscillatory
fitting functions whose sole role is to soak up the statistical fluctuations on
amplitudes known to be zero. Note that we include the constant function as it
has an appreciable expansion on modes of $\Sigma$ with large eigenvalues.

In our fitting procedure we keep about six modes but check the the results are
stable for all reasonable choices.

\section{Numerical Results }

Our numerical results were obtained at a couplings $g=1.5$ on the $32\times 32$
and $g=1.75$ on the $64\times 64$
lattices respectively. Measurements were taken every two updates after
equilibriation
at values $\Delta\tau^2=0.04,~0.06~\&~0.08$ with a squared fourier acceleration
mass of $M^2=.02~$ .
The total number of iterations used was $48\times 10^3$, with $32\times 10^3$
at $\Delta\tau^2=0.04$,
and $8\times 10^3$ at both of $\Delta\tau^2=0.06~\&~0.08~$.  We employed a
higher number
of iterations at the lowest value of
$\Delta\tau^2$ because of its importance in the extrapolation procedure to
$\Delta\tau^2=0~$.

The results are accumulated according to a standard re-binning procedure in
which means, mean squares and
cross correlators are compiled from individual results then from the averages
of pairs, quadruples etc.
By computing the variance of an observable from a sequence of bins and noting
the point at which the
it stabilises we can estimate
a value for the Langevin correlation time of the observables in question. The
value we quote
$\tau_{\chi}$, is that appropriate to the susceptibility.
By applying this  approach with and without operator improvement we can detect
the effect of
operator improvement on the Langevin decorrelation time.

In Table 1 we exhibit the results of our procedure for the
susceptibility $\chi$, and the Langevin decorrelation
time $\tau_{\chi}$ for lattice sizes of $32\times 32$ and $64\times 64$.
In Table 2 we show the results for the correlation length $\xi$~.
The values of chi$^2$ per degree of freedom are for the
$\Delta\tau^2$-extrapolation.
The results for $\chi$ and $\xi$ compare quite well with the corresponding
results
obtained by Meyer who used a multi-grid method.
We have comparable or smaller statistical errors for a smaller number of
iterations of the
update.

Fig 1 shows a comparison between extrapolations for the susceptibility on
the $64\times 64$-lattice with and without the operator improvement. The value
of
$\mbox{chi}^2$ for these extrapolations were very much less than unity
indicating
the assumption of $O(\Delta\tau^2)$ for the systematic errors was well
verified.
It is clear from Fig. 1 that the operator improvement also leads to a reduction
in the
the {\em systematic} error, a fact which helps to improve the extrapolation
procedure by reducing the obliqueness of the angle at which the extrapolation
line hits the
$\Delta\tau^2=0$ axis. The reason may be inferred from the following analysis.
If we accept
that our choice for $W(z)=1+bz$ is an approximation to the choice
$W(z)=\exp\{bz\}$ then we see
that our operator improvement is an approximation to
$f'(S)=\exp\{bH^{\dag}f(S)\}$~. The
simulation results in a probability distribution $P_{\mbox{sim}}(S)$ which is
in error by
$O(\Delta\tau^2)$. The simulation average of the improved operator is then
\be
\la f'(S)\ra_{\mbox{sim}}=\int d\mu(S)P_{\mbox{sim}}(S)e^{bH^{\dag}}f(S)~~.
\ee
Integration by parts yields
\be
\la f'(S)\ra_{\mbox{sim}}=\int
d\mu(S)\left(e^{bH}P_{\mbox{sim}}(S)\right)f(S)~~.
\ee
The effective probability distribution is therefore
$P_{\mbox{eff}}(S)=\exp\{bH\}P_{\mbox{sim}}(S)$~.
Clearly $P_{\mbox{eff}}(S)\rightarrow P_0(S)$ as $b\rightarrow \infty$~.
Our variance reduction method does yield positive values of $b$ because the
operator
improvement correlates strongly and negatively with the original estimator
(eq(\ref{sig})). That this must be the case follows directly from the
non-positive
nature of $H^{\dag}$. Consequently, it is plausible that we can also expect a
reduction in
the systematic error for the improved operator.
We find that this is indeed the case.
The negative correlation is illustrated in Fig. 2 where we have plotted the
improved operator
results for the various values of
$\Delta \tau^2$ against $b$. The bands represent the the result $\pm$ the
statistical error
and not only exhibit the expected narrowing corresponding to minimum variance
but also
show a tendency to converge around a region encompassing the true result thus
indicating that the operator
improvement compensates in a consistent way for the systematic error. This
latter
compensation is not perfect of course but is not required to be and might be
further improved
by including additional $(H^{\dag})^2$ terms in the operator improvement.

In Figs. $3~\&~4$ we show the extrapolated correlation function $G_t$ together
with the fit obtained by
the above procedure for $32\times 32$ and $64\times 64$ lattices. In line with
our general procedure
these fits were obtained with the omission of the first three $t$-values. The
values for the
chi$^2$ per degree of freedom were approximately unity. The ommission of
further $t$-values
changed the fitted parameters only by small amounts well within the statistical
errors.

{}From the results in Table 1 it is clear that operator
improvement drastically reduces the Langevin correlation time. In the case of
the $32\times32$-lattice
the reduction is by a factor of 3.5 from 14 time-steps down to 4 and for the
$64\times64$-lattice by a factor
of 4 from 16 to 4~. These speeded up decorrelations can be plausibly
interpreted as the
origin of the reduced variances that we achieved after operator improvement.  A
comparison of our
results with those of ref \cite{HasMey1} suggests that for a given number of
iterations our
statistical errors are better by a factor of a little more than two. Translated
into numbers of iterations
this becomes a factor of approximately four or five.
Allowing for differences in computers we estimate that our CPU update time
is approximately six times greater than that of refs \cite{HasMey1,HasMey2}.
The net
result is that we achieve, for $N=3$ on the sizes
of lattice we have so far been able to deal with, a given error in a comparable
real time notwithstanding
our need to accumulate data at different values of $\Delta\tau$ in order to
perform the
extrapolation needed to eliminate the systematic error.

A comparison of methods needs a detailed numerical study. Some general points
can be made however.
The time consuming parts of the update step that we use are the exponentiation
of the generators
to compute $SU(3)$ group elements and the Fourier transform procedure required
for the implementation
of the Fourier accelerated update. The exponentiation of the generators to
produce finite group elements
is at the centre of the various updating methods. Hasenbusch and Meyer
\cite{HasMey1,HasMey2}
simplify this problem by organising the algorithm so that only diagonal
generators need to be
exponentiated. This leads to a dependence on $N$ for their algorithm that is
$O(N)$~. In our
case for $SU(N)\times SU(N)/SU(N)$ the dependence is necessarily $O(N^3)$~. For
$CP(N)$,
which we have not investigated in this paper, it would be $O(N^2)$~.
Superficially then it
would seem that our Langevin scheme would lose rapidly to the updating
technique used by Hasenbusch
and Meyer for larger $N$ values. However the outcome of any comparison depends
also on the efficacy
of the update in covering the configuration space of the spins. We have seen
that per update
our method augmented with operator improvement does a relatively better job in
producing
independent configurations.
There are grounds to believe that this relative advantage will increase with
$N$ and that the method will remain competitive with that of Hasenbusch and
Meyer
\cite{HasMey1,HasMey2}.
Only a numerical trial however can properly settle this issue.

We are presently applying the Langevin method with operator improvement to
larger lattices
in order to test for the occurence of critical slowing down. A  vitally
important aspect of the problem
that we have not fully addressed in this paper.

\section*{Acknowledgements}
RRH would like to thank the Theory Division CERN for their hospitality
during a sabbatical year when much of this work was carried out, and he
would also like to thank U. Wolff for useful discussions.
\pagebreak

\section*{Tables}
{}~~~~~~~~~~~~~~~~~~~~~~~~~~~~~~~~~~
\vskip 5 truemm
{\bf Table 1}
\vskip 3 truemm

\begin{tabular}{|l|l|l|c|c|r|}\hline
$g$&$L$&$b_{\chi}$&$\chi$&chi$^2$&$\tau_{\chi}$  \\ \hline\hline
1.5&32&0.0&59.0(7)&0.02&14 \\  \hline
1.5&32&0.6&57.7(4)&0.52&4 \\ \hline
1.75&64&0.0&298.8(4.2)&0.76&11 \\ \hline
1.75&64&0.8&301.5(2.8)&0.003&3 \\ \hline
\end{tabular}

\vskip 5 truemm
{\bf Table 2}
\vskip 3 truemm

\begin{tabular}{|l|l|l|c|r|}\hline
$g$&$L$&$b_{\xi}$&$\xi$&chi$^2$   \\ \hline\hline
1.5&32&0.0&3.04(3)&0.94 \\  \hline
1.5&32&0.2&3.05(2)&1.2 \\ \hline
1.75&64&0.0&8.39(09)&1.5 \\ \hline
1.75&64&0.2&8.41(07)&1.5 \\ \hline
\end{tabular}

\section*{Table Captions}
\begin{itemize}
\item []Table 1. List of results for the susceptibility $(\chi)$ on
two lattices with and without operator improvment. The chi$^2$ per degree of
freedom is shown for the $\Delta\tau^2$-fit.

\item []Table 2. List of results for the correlation length $(\xi)$ on
two lattices with and without operator improvment. The chi$^2$ per degree of
freedom is shown for the $\Delta\tau^2$-fit.

\end{itemize}

\section*{Figure Captions}

\vskip 5 truemm
\begin{itemize}
\item[] Fig. 1. Plot of the susceptibility against $\Delta\tau^2$~ on a
$64\times 64$-lattice.
The upper and lower curves show the results with and without operator
improvement.

\item[] Fig. 2. Plot of susceptibility on a $64\times 64$-lattice against the
improvement
parameter $b$ for $\Delta\tau^2=0.04~$(continuous line), 0.06~(light
dashes)$~\&~0.08$~(heavy dashes). The bands indicate the statistical error.

\item[] Fig. 3. The correlation function on a $32\times 32$-lattice for
coupling $g=1.5$~.

\item[] Fig. 4. The correlation function on a $64\times 64$-lattice for
coupling $g=1.75$~.

\end{itemize}

\pagebreak

\end{document}